\documentstyle[11pt]{article}

\begin{document}

\vspace*{26mm}

{\Large
\noindent
\bf The Venus Tablet and Refraction}\\[3mm]

{\large V.G.Gurzadyan}\\[6mm]
{University  of  Rome ``La Sapienza", Italy  and   Yerevan  Physics
Institute, Armenia}\\[5mm]

(Published in {\it Akkadica},  v. 124 (2003), pp. 13-17.)

\section{Introduction}

For several decades the solution to the problem of the ancient Babylonian chronology was anchored to the belief in the existence of 56/64 year cycles in the Venus Tablet of Ammisaduqa.  That belief had determined the choice of the possible chronologies, reducing them to those known as High, Middle and Low chronologies. The earlier chronological proposals had assumed a link between a fixed lunar calendar and the 56/64 year cycles of the visibility of Venus.  One novelty of the astronomical background of the "New" or "Ultra-low" chronology advocated in recent years was the understanding that the 56/64 year cycles cannot in fact be accurately recognized in the Venus Tablet (Gasche et al. 1998; Gurzadyan 2000).  This implied a definite change in the framework of chronological debate, since the 56/64 year cycles could no longer serve as a basis for argumentation.

We also addressed the possibility of astronomical arguments based on ancient observations, namely, the Enuma Anu Enlil Tablets 20 and 21. The approach depended upon extracting and analysing the astronomically most significant descriptors which interpreters of the tablets agreed had survived in the descriptions of the eclipses.  The analysis was based on the data preserved concerning the exit position of the darkening of the lunar disk and the time recorded.  These were relatively the most reliable and astronomically most informative since the reality of observations is that, the end of an astronomical event is generally better recorded than its beginning. The chronology of the Third Dynasty of Ur indicated that two lunar eclipses should have been separated by a period of 41-44 years.  Therefore the search centered on two eclipses which could be related relatively in time.   Along with the necessary gap, the exit positions and the watchtime information determined the choice of the eclipses of 1954 BC and 1912 BC as the most probable candidates for the Ur III events.  
This strategy, i.e. exploiting the known lunar eclipses from the omina together with the 8-year Venus-cycle from the Venus Tablet, revealed its value in the creation of a new chronology, suggesting that the fall of Babylon corresponded to 1499 BC (rather than 1531 or 1595 as posited by the Low and Middle Chronologies).  This Ultra-Low chronology is also compatible with such independent data as pottery sequences, Assyrian king lists, and glyptic evidence.

\section{Refraction}

Here, we continue the discussion of astronomical effects which can influence the interpretation of the omens and even the choice of the chronology.  Particularly, we will concentrate on the role of refraction in observations near the horizon.  Observations near the horizon are the key ones for the Venus Tablet since the latter includes the dates of the first and last visibilities of the Venus.
The consideration of refraction corrections had already led to the refutation of the megalithic lunar observatory idea (Schaefer and Liller 1990), due to the fact that refraction exhibits variations which are more than a magnitude higher than the values adopted previously by archaeoastronomers.  Such variations are also difficult to predict based on the atmosphere models in the widely used programs of evaluation of refraction (Garfinkel 1967).

We will therefore inquire whether the role of refraction can be of such profound importance for the Venus Tablet interpretation as it was for megalithic astronomy.
The refraction near the horizon $R_0$ is defined as 
$$
R_0 = Z-90^{\circ}-D,
$$
where $Z$ is the angular distance of the planet or of the upper limb of the Sun from the zenith, $D$ is the correcting factor due to the altitude of the observer.
The quantitative evaluation of refraction in a given site can be performed by several methods, for example, via the difference in the timing of the actual contact of the planet with the horizon and theoretical tables, or the calculation of the index via the known profile of the temperature using the corresponding formulae, or the software by Schaefer (1989).

The actual observations performed by Schaefer and Liller (1990) at two sites range for R0 from $0^{\circ}.234$ to $1^{\circ}.678$, with the conclusion of reasonable variation of $R_0$ for an unknown site about $0^{\circ}.64$.

In March, 2001 we had a brief opportunity to perform observations from the Northern Palace in Babylon, particularly we checked that the apparent horizon in Babylon is close to the mathematical horizon due to the flatness of the site; nor can west-east asymmetry be observed.  The altitude of our location was about h=20m above the level of Euphrates, which guarantees the smallness of the factor
$$
D=arccos(1/(1+h/R_{Earth})).
$$
Hence, one could expect that the estimations of refraction variation by Schaefer and Liller (1990) have to be valid for this site also.  Although we did not have occasion for detailed observations, either to trace the systematic effects or to collect considerable statistics, our estimation on the site of $R_0$ based on the actual and theoretical timing, about $0^{\circ}.6-0^{\circ}.8$, was in accordance with those of Schaefer and Liller (1990); inaccuracies due to the variation of climatic conditions since the Old Babylonian period would hardly exceed that scatter.

Turning now to the Venus Tablet and the Venus-cycles, we recall that five synodic periods of Venus are almost equal to eight sidereal years, namely, after 8 years the coordinates of Venus are repeated with a shift of almost $2^{\circ}.5$ which yields roughly a 4 day shift in the dating.  Adopting the value of Schaefer and Liller (1990) as a typical refraction variation, $0^{\circ}.64$, we see that it will introduce inaccuracies in the dating of the rise or set of Venus ranging from 1 to 4 or 5 days, i.e. in the dates of the first and last visibilities.  This multiple-day inaccuracy is crucial for the link between the Venus Tablet data and the lunar calendar, and hence for chronological purposes.

According to the Omens 1 and 57 of the Venus Tablet, Venus had its last visibility on the 15th day and first visibility on the 18th day of month XI of the first year of Ammisaduqa, i.e. it was invisible for 3 days (Reiner and Pingree 1975).  The comparison with modern calculations (Van der Waerden 1974), links this fact with the end of March of 1701 BC, which then, together with the assumed 56/64 periodicity became the basis of the fixing of 1702 BC, 1646 BC resp. 1638 BC, and 1582 BC, as the proposed dates for the first years of Ammisaduqa, i.e. the High, Middle and Low chronologies.  As a result, subsequent discussion concentrated on the choice of the most probable chronology among these.  Van der Waerden (1974), particularly, found the Short or Low the highest probable, while Huber (1987) advocated the High.  
However, as mentioned above, the elimination of the chronological anchors of the 56/64 year cycle liberated the search from those bounds.  We can also now see further evidence for the absence of the former alleged bounds, due to the intrinsic inaccuracy introduced by refraction.  The signature of refraction variations have to survive in the original core of the Venus Tablet, precisely because - in contrast to, say, weather conditions - it will have a more regular character.  Moreover, one can consider it as a criterion for revealing the genuine part of the Tablet, i.e. by means of the consideration of refraction corrections on the dates of Venus visibilities given in the Tablet.  Dates which fail to satisfy such regular distortions must be identified either as (a) not reflecting actual observations or (b) later addendums.  More detailed studies of refraction in Babylonian conditions, along with the deeper analysis of the noise in the Venus Tablet, will be required for that purpose.

\section{Solar eclipses}

Let us also briefly discuss recent attempts to link the reference to a solar eclipse related to the birth of Shamshi-Adad to the various chronological schemes.  Michel and Rocher (1997-2000) proposed the eclipse of November 19, 1795 BC.  Michel (2002) subsequently proposed  the eclipse of June 24, 1833 BC.  Warburton (2002) has mentioned the eclipses of April 2, 1847 BC and October 8, 1764 BC.

We will particularly outline the ambiguities arising at such attempts for particular eclipses.  Among the eclipses discussed, the solar eclipse of October 8, 1764 BC would have been observed at Mari, as well as in Babylon and in Assur with a close magnitude and a shift in time of only a few minutes, as an annular, non-central eclipse.  The peculiarity of this historical eclipse is that the location of the shadow at its maximal phases makes the fact of the appearance of the eclipse in Mari independent of the time difference Delta T= ET-UT, i.e. on the principal uncertainty affecting most ancient eclipses due to the unknown rate of the variation of the Earth's rotation period determined by tidal and non-tidal effects.  For that eclipse only the magnitude of the maximal phase depends on that uncertainty, but not in a crucial way.  
The solar eclipse of April 2, 1847 BC also, would have been observed both in Mari and Assur, however its appearance, i.e. the magnitude and the duration, is affected more by the uncertainty in Delta T.  The total eclipse of June 24, 1833 BC might have been observed in Assur, but only with a magnitude of about 0.4, quite low to be recorded as a presumably major historical event.  The latter estimation corresponds to ET-UT=49010.8 sec, and at present there is no way for more precise theoretical estimations.  Similarly, the total eclipse of November 19, 1795 BC would have observed as a partial eclipse with relatively higher maximum magnitude in Assur than in Mari (though both lower than those of the eclipse of October 8, 1764 BC), however the absolute values of the magnitudes again remain inaccurate.

\section{Conclusion}

The record of a single solar eclipse of Old Babylonian date, without description of details and/or unambiguous historical links, can hardly act as a reliable anchor.  The relevance depends upon the ability to link the eclipse to another anchored astronomical event or to another similarly anchored chronological framework. Given the dates for the life and death of Shamshi-Adad and the various proposed chronologies, one could argue that the eclipse of 1847 BC is compatible with the Middle Chronology, or that of 1795 BC with the Low Chronology, or that of 1764 BC with the Ultra-Low Chronology.  That of 1833 BC is not compatible with any other existing chronological framework.  However, all of these potentially relevant eclipses suffer from the difficulty of trying to base chronologies on a single eclipse without additional information.  The selection of a date cannot alone resolve the chronological debate unless the astronomical data are linked to the ancient sources.
Refraction acts as an additional uncertainty, hitherto neglected in the interpretation of the Venus Tablet, in the fitting the data not only to fixed lunar calendar, but also to the moving lunar calendar.  The only robust signal which can be extracted from the Venus Tablet, therefore, remains the 8-year cycle as argued in (Gasche et al. 1998; Gurzadyan 2000).

\renewcommand{\baselinestretch}{1}
\section*{\it References}

\begin{description}
\item[ ]\hspace{-1mm} GARFINKEL B., 1967: "Astronomical Refraction in a Polytropic Atmosphere", {\it Astronomical Journal} 72, 235-254.    \\[-7mm]
\item[ ]\hspace{-1mm} Gasche et al. = GASCHE, H., ARMSTRONG, J.A., COLE, S.W., GURZADYAN, V.G., 1998: {\it Dating the Fall of Babylon. A Reappraisal of Second-Millennium Chronology} (= MHEM 4), Ghent, Chicago.  \\[-7mm]
\item[ ]\hspace{-1mm}  GURZADYAN, V.G., 2000: "On the Astronomical Records and Babylonian Chronology", {\it Just in Time. Proceedings of the International Colloquium on Ancient Near Eastern Chronology (2nd Millennium BC)}. Ghent 7-9 July 2000 (= {\it Akkadica 119-120)}, Bruxelles, 177-186.    \\[-7mm]
\item[ ]\hspace{-1mm} HUBER, P.J., 1987: "Astronomical Evidence for the Long and Against the Middle and Short Chronologies" in ASTRÖM, P. (éd.), {\it High, Middle or Low ?} Acts of an International Colloquium on Absolute Chronology Held at the University of Gothenburg 20th-22nd August 1987. Part 1 (= Studies in Mediterranean Archaeology and Literature. Pocket-book 56), Gothenburg, 5-17.
  \\[-7mm]
\item[]\hspace{-1mm} MICHEL, C., 2002: "Nouvelles données pour la chronologie du IIe millénair", {\it NABU} 20).\\[-7mm]
\item[ ]\hspace{-1mm} MICHEL, C., ROCHER, P., 1997-2000: "La chronologie du IIe millénaire revue à l'ombre d'une éclipse de soleil", {\it JEOL} 35-36, 111-126.
  \\[-7mm]
\item[ ]\hspace{-1mm} REINER, E., PINGREE, D., 1975: {\it Babylonian Planetary Omens. Part 1: The Venus Tablet of Ammisaduqa} (= BiMes 2/1), Malibu.
  \\[-7mm]
\item[ ]\hspace{-1mm} SCHAEFER, B.E., LILLER, W., 1990: "Refraction Near the Horizon", {\it Publications of the Astronomical Society of the Pacific} 102, 796-805. \\[-7mm]
\item[ ]\hspace{-1mm} SCHAEFER, B.E., 1989: "Refraction by Earth's Atmosphere", {\it Sky and Telescope} 77, 311-313.
  \\[-7mm]
\item[ ]\hspace{-1mm} VAN DER WAERDEN, B.L., 1974: {\it Science Awakening II. The Birth of Astronomy}, Groningen.
  \\[-7mm]
\item[]\hspace{-1mm} WARBURTON, D.A., 2002: "Eclipses, Venus-Cycles \& Chronology", {\it Akkadica} 123, 108-114.
  \\

\end{description}

\end{document}